\documentclass{aa}
\usepackage{txfonts}
\usepackage{color}

\newcommand{\onvire}[1]{}

\usepackage{color}
\usepackage{epsfig}
\usepackage{amssymb}
\usepackage{graphicx,natbib}
\definecolor{grey}{cmyk}{0.,0.,0.,0.5}

\begin{document}

\title{Spallation dominated propagation of Heavy Cosmic Rays and the 
Local Interstellar Medium (LISM)}
\author{C. Combet\inst{1,2}, D. Maurin\inst{3}, J. Donnelly\inst{1}, 
L. O'C Drury\inst{1}, E. Vangioni-Flam\inst{4}}

          \authorrunning{Combet et al.}
          \titlerunning{Spallation dominated regime}
          \institute{Dublin Institute for Advanced Studies (DIAS), 
	5 Merrion Square, Dublin 2, Ireland
	\and Laboratoire de l'Univers et de ses Th\'eories (LUTh), B\^at.18, 
	5 pl. J. Janssen, 92195 Meudon, France
	\and Service d'Astrophysique (SAp), CEA, 
        Orme des Merisiers, 91191 Gif-sur-Yvette, France
	\and Institut d'Astrophysique de Paris (IAP), 98 bis Bd
        Arago, 75014 Paris, France 
	}

\offprints{C. Combet, \email{combet@cp.dias.ie}}

\date{Received \today; accepted}

\abstract{Measurements of ultra heavy nuclei at GeV/n energies in the galactic
cosmic radiation address the question of the sources (nucleosynthetic
s- and r-processes). As such, the determination of CR source abundances is a
promising way to discriminate between existing nucleosynthesis models.
For primary species (nuclei present and accelerated at sources), 
it is generally assumed that the relative propagated
abundances, if they are close in mass, are not too different from their
relative source abundances. Besides, the range
of the correction factor associated to propagation has been estimated in 
weighted slab models only. Heavy CRs that are detected near Earth were 
accelerated from regions that are closer to us than were the light 
nuclei. Hence, the geometry of sources in the Solar neighbourhood, 
and as equally important,
the geometry of gas in the same region, 
must be taken into account. In this paper, a
two zone diffusion model is used, and as was previously investigated
for radioactive species, we report here on
the impact of the local interstellar medium (LISM) feature
(under-dense medium over a scale $\sim100$~pc) on primary and
secondary stable nuclei propagated abundances. Going down to Fe nuclei, 
the connection between heavy and light abundances is also inspected.
A general trend is found that decreases the UHCR source abundances relative to
the HCR ones. This could have an impact on the level of r-process 
required to reproduce the data.
}

\maketitle


\section{Introduction}

Beyond the iron peak, the flux of Cosmic Ray nuclei drops by several
orders of magnitude. The UHCRs require peculiar environments to be
nucleosynthesised and two distinct processes are involved:
very generally, either the neutron flux in the medium is low so that
the neutronic capture rate is less than the $\beta$-decay rate of nuclei, or the
flux is high (capture rate much greater than $\beta$-decay rate); these two
situations are commonly referred to as s- (slow) and r-process (rapid). 
The situation is obviously more complex --~ 
see~\citet{1994ARA&A..32..153M}
for an illuminating review --. One particular issue
is the determination of the relative contributions of s- and r-process nucleosynthesis
required to explain the data, inferring the corresponding source abundances
from the elemental and isotopic abundances.

Based on model
predictions~\citep{1994ARA&A..32..153M} as well as on
some analysis of CR data~\citep{1983ApJ...264..324B}, it is shown, 
for example, 
that for $Z\geq 89$, the entire Solar System abundances must be attributed to
the r-process. On the opposite, the latter does not
contribute to $Z\leq 40$~\citep{1981ApJ...247L.115B}.
In-between, most of the elements exist as a mixture of r- and s- contributions.
Their origin (e.g., massive stars, explosive nucleosynthesis) and exact abundances
are still debated~\citep{1999A&A...342..881G,2004oee..symp...27C} and CRs
could provide some answers. 
However, to determine the CR source abundances, the
fluxes measured near Earth must be propagated back through
the Galaxy, which is not a straightforward task.

We shall not discuss here the important question of the
time elapsed since nucleosynthesis and propagation which is studied
through radioactive heavy elements \citep{2002SSRv..100..277T} and which is
another part of the puzzle, nor shall we use any propagation network
to study all the abundances \citep{1984ApJS...56..369L}. We instead focus 
on propagation. The goal of
this paper is to demonstrate the importance of the local CR-source
distribution and in particular to examine the effect of the local gas distribution 
on calculations of source abundances from data.
As a matter of fact, whereas the truncation of path lengths for UHCRs
has been discussed (through the weighted slab approach) and recognised
as having a large impact on the propagation of UHCRs
\citep{1993ApJ...403..644C,1996ApJ...470.1218W}, 
we illustrate here
how this truncation is realised as a deficit in nearby sources in the
two zone diffusion model. The low density gas
in the local interstellar medium (LISM) is also of particular importance.
 It is shown that this under-dense region leads to results
different from both the standard two zone diffusion model (which is equivalent
to a simple leaky box model) and the truncated
weighted slab.

The main results of this study are: i) around a few GeV/n, 
the relative abundance of $Z>60$ nuclei
to the lighter ones is decreased when the LISM features are taken 
into account --~the strength of the effect is inversely related to the value of
the diffusion coefficient (i.e. to the energy) and is directly related to 
the destruction cross section values (i.e. to $Z$)~--; ii) the shape of the spectra
at low energies is sensitive to the value of the diffusion coefficient, 
providing a way to determine the latter as it breaks the propagation 
parameter degeneracy \citep{2002A&A...394.1039M} 
observed in standard diffusion model (i.e., with constant gas and source
disk distribution); 
iii) in the UHCR context of
relative abundance determination (Pb/Pt and Actinides/Pt),
a gas sub-density has a major effect for mixed species (species
with both primary and secondary contributions) only.

In~$\S 2$, we give some simple arguments which allow us to understand
why the local sub-density is expected to have a major influence on heavy CRs
at GeV/n energies. In~$\S 3$, the model taking into account
the sub-density as a circular hole is exposed. Profiles in the LISM
cavity as well as spectra for primary and secondary-like nuclei under
various source geometry assumptions
are compared to those obtained in a standard ``no-hole'' diffusion model.
The correction factors used to determine the source abundances are then derived
and consequences for the data are discussed in~$\S 4$.
We then conclude and comment on further developments demanded by UHCRs
propagation.


\section{ Why a local sub-density may affect the propagation}
\label{Sec:}

It has been previously recognised than very heavy
nuclei have a peculiar propagation history compared to light nuclei.
This is related to their large destruction cross sections which make
them have very short path lengths. Regarding this
extreme sensitivity to nuclear destruction, an UHCR 
breaks-up more often than a lighter one and thus propagates
for a shorter distance.
\cite{1972PhysRevD...5..307} emphasized that the
strength of the source and its location in space and time may be
a dominant local contribution or merely adds to the general background
depending on whether this source is close or far. These authors use a very simple
model in which propagation starts from a single source in space and time.
They conclude that $A>81$ elements cannot be explained by any source older
than $10^6$~yr. In the present paper, the steady-state approximation is
however assumed but the spatial discreteness hypothesis is relaxed to
some extent. This is a first step away from a leaky box model towards a
more realistic description. The framework is a two zone diffusion model
(thin disk and halo) which has been used by many authors for
light nuclei, e.g.~\citet{2001ApJ...555..585M}. In particular, such a model has
distinct
features compared to leaky boxes. The interstellar matter density
is now located in a thin disk with CRs spending most of their time
in the diffusive halo. The characteristic times of the various
processes in competition during propagation (see below) 
can be extracted~\citep{2003A&A...402..971T}. As it
is shown below, the fact that the typical distances
travelled by heavy or UHCRs before their destruction could be
as little as a few hundreds of pc naturally means that 
LISM properties are an important ingredient
for propagation.

\subsection{Propagation characteristic times {\em vs} A}
\label{Sec:Charac_times}
During the propagation of nuclei through the 
Galaxy, several processes affect their journeys. All of these
have specific time/length scales which characterise the
importance of the considered process relative to the others.
The nuclei can undergo energy losses
of different forms, interact with atoms of the ISM and
be destroyed in a spallation process or
escape from the boundaries of the diffusive volume (i.e.
the galactic halo) because of the combined influences of 
diffusion and convection.

In a leaky box model, the density of gas in the box is constant
so that the link between characteristic lengths and times
is simple and the comparison to the escape length
straightforward. In reality,
spallation and energy losses occur when a nucleus crosses
the galactic disk. Thus, the characteristic times (or lengths) for those
processes are directly linked with the number of
disk-crossings~\citep{2003A&A...402..971T} which should be compared
to the diffusive and convective escape times.

\subsubsection{Spallation dominates over energy losses}
\label{tpssimples}
Whether in a leaky box model or a more sophisticated model as used here,
spallations and losses both occur when gas is traversed. We
assume its density to be $n_{\rm{ISM}}=1$~cm$^{-3}$.
The spallation rate is given by $\Gamma_{\rm sp}=n_{\rm{ISM}}\sigma v$, where $v$
is the velocity of the nucleus and $\sigma$ is the reaction cross 
section. For this study, the \citet{1983ApJS...51..271L} cross sections
are accurate enough ($E_{k/n}$ is the kinetic energy per nucleon):
\[
\sigma=\sigma_\infty[1-0.62 \exp(-E_{k/n}/200)\sin(10.9 E_{k/n}^{-0.28})]\,,
\]
\[
\sigma_\infty(\rm{E_{k/n}>2 GeV/n})=45 A^{0.7}[1+0.016 \sin(5.3-2.63\log(A))]\,.
\]
The loss rate is given by 
$\Gamma_{\rm{loss}}=\left(dE/dt\right)_{\rm{loss}}/E_k$, where $E_k$ is 
the kinetic energy of the nucleus, and the Coulomb and ionisation energy
losses are taken into account \citep{1994A&A...286..983M,1998ApJ...509..212S}
assuming the ionised fraction of the ISM to be 0.033.
In both cases, $dE/dt$ is proportional to $Z^2$, with $Z$ the charge
of the nucleus.

At a given $E_{k}/n$, $\Gamma_{\rm sp}$ scales with 
the atomic mass $A$ as $A^{0.7}$ ($\log A$ does not 
induce strong variations). Assuming $A=2 Z$, which is not the case
for heavy nuclei ($A>2 Z$), ionisation and Coulomb loss
rates also increase with A but following $\Gamma_\mathrm{ion/coul}\propto A$.
A comparison of these two rates yields at 1~GeV/n 
\[
\Gamma_\mathrm{sp}\sim 10 \times(A/10)^{-0.3}\;\Gamma_\mathrm{ion},
\]
\[
\Gamma_\mathrm{sp}\sim 60 \times(A/10)^{-0.3}\;\Gamma_\mathrm{coul}.
\]
This means that the effect of spallation is always dominant 
over energy losses (at higher energy the effect is even stronger).
The latter are thus discarded in the rest of this paper as it
mainly focuses on qualitative effects.
Note also that the cross sections have not been corrected for the effects of
decay which can be significant for the propagation of heavy nuclei. These effects,
as well as energy losses and reacceleration, will be properly taken into account
in a separate paper where $\beta$-decay and electronic capture decay will be re-examined
in the context of the present diffusion models.

\subsubsection{Time scales of diffusion, convection and spallation}
\label{time_scales}
Once established that spallations are dominant relative to
the energy losses, we compare the former to the convection and
diffusion processes taking into account
the number of disk-crossings in the two zone diffusion model.
For further justification and details the reader is referred
to~\citet{2003A&A...402..971T}.

It is well known that $\tau_{\rm{diff}}=L^2/K(E)$
is the characteristic escape time needed by a nucleus to leave
a diffusive volume of size $L$ for a diffusion
coefficient $K(E)$.
The presence of a galactic wind (velocity $V_c$),
assumed to be constant and perpendicular to the galactic plane,
induces a general convective motion throughout the Galaxy.
Considering that the 
diffusive process is still present, the boundary to be 
taken into account is not the halo size but the distance for which 
the diffusion cannot compete anymore with the convection and
brings the nucleus towards the Earth. 
With this consideration, one can define the typical time scale
for the convection escape process as $\tau_{\rm{wind}}=2K(E)/V_c^2$.

Spallation only occurs within the disk of total thickness $2h$.
The typical length associated with the spallation 
process in a diffusive propagation is $r_{\rm{sp}}=K(E)/(h\Gamma_{\rm{sp}})$.
The time scale, assuming a purely diffusive transport is then
$\tau_{\rm{sp}}=r_{\rm{sp}}^2/K(E)=K(E)/(h\Gamma_{\rm{sp}})^2$.
\begin{figure}
\begin{center}
\includegraphics[bb=12 50  730 580, width=8. cm]{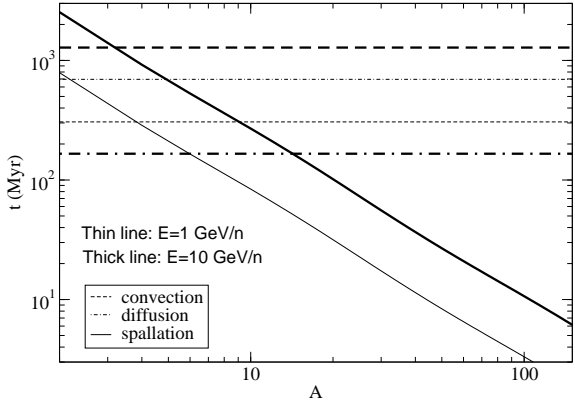}
\caption{Characteristic times of diffusive/convective escape and
spallation as a function of the atomic mass. Typical 
values $K_0=0.0112$~kpc$^2$Myr$^{-1}$ and $L_{\rm{halo}}=4$~kpc
were considered. The times have been calculated at kinetic
energies per nucleon of 1~GeV/n and 10~GeV/n in the
two zone diffusion model.\label{t_vs_A2}}
\end{center}
\end{figure}

The characteristic times for diffusion, convection and spallation
are plotted on Fig.~\ref{t_vs_A2} as a function of the atomic mass
of the nucleus $A$ at two different kinetic energies per nucleon
(1~GeV/n and 10~GeV/n). 
As we assume $K(E)=K_0 \beta (\mathcal R/1\rm{GeV})^\delta$ 
(where $\mathcal R$ is the rigidity), the convection and diffusion times are
constant with A. 
At 1 GeV/n, it appears that spallation is almost
always dominant, except for the very light nuclei where both convection
and diffusion compete. At 10 GeV/n, spallation is less
dominant but remains the major process for the heavier
nuclei. A naive (and false) interpretation of these figures is that
a heavy nucleus never escapes from the Galaxy. 
Actually, many paths lead from the source to
the halo boundary and these numbers only indicate the typical time
a CR reaching our position can have travelled before being destroyed.
For convection and diffusion, $\tau_{\rm{diff}}$ and $\tau_{\rm{wind}}$ define
an exponential cutoff whereas the cutoff is smoother
for spallation~\citep{2003A&A...402..971T}.

Table~\ref{r_spal} contains the typical spallation distances for different
atomic masses (10, 50, 100 and 200) and a set of three diffusion coefficients
(two extreme values and one median) which fit the B/C 
ratio when combined with the appropriate choice for the other propagation
parameters \citep{2001ApJ...555..585M,2002A&A...394.1039M}.
For A=100 and A=200 and for the median value of the 
diffusion coefficient, the spallation length scales
have the same order of magnitude as the size of the observed
local gas sub-density (see Sect.~\ref{Sec:Ingredients}).
This sub-density precludes the creation of secondary species 
in the local bubble and the average density crossed by a primary
or a secondary nucleus during its journey is smaller than when
this sub-density is not taken into account.
It is expected that heavier nuclei will be more sensitive
than lighter nuclei to this feature.
\begin{table}[ht]
\begin{center}
\begin{tabular}{c c c c c}
\hline
\hline
A &Group &  & $K_0$ (kpc$^2$~Myr$^{-1}$) & \\
    &   & 0.0016 & 0.0112 & 0.0765 \\ \hline
10 & LiBeB/C & 210 pc  & 1.25 kpc & 6.38  kpc  \\
50 & Sub-Fe/Fe & 70 pc  & 390 kpc &  2.01 kpc  \\
100 & Z=44-48 & 40 pc  & 250  pc & 1.27   kpc   \\
200 & Actinides & 30 pc  &  150 pc & 780 pc \\
\hline
\end{tabular} 
\caption{Typical length scales of the spallation process for different
nuclei and diffusion coefficients at 1 GeV/n. \label{r_spal}}
\end{center}
\end{table}

The smaller the spallation length scale, the greater the effect
of the sub-density is expected to be and this suggests
that the local bubble must be considered to treat 
the heavier species. This reasoning is valid
only if the cosmic rays undergo the same diffusive process in
the LISM than in the rest of the Galaxy. There is at least two
indications supporting this assumption:
\begin{itemize}
  \item The local bubble and the galactic halo share some common
properties. Hence, it seems reasonnable that the values of
the diffusion coefficient in both regions are not drastically
different. As we use only one coefficient for the disk
and the halo, the latter can be seen as an effective value that
also applies for the local bubble.
  \item The second clue comes from the analysis of secondary 
radioactive species which can travel for a few hundreds of pc
only before decaying. \citet{1998A&A...337..859P} were able to derive the
corresponding 'local' diffusion coefficient modelling
the LISM as three shells of gas with various densities.
They found a larger value at low energy than the standard
one found from B/C analysis. This again points towards a diffusive
transport of CRs in the local bubble.
\end{itemize}


\section{Two zone model with a hole}
\label{Sec:Ingredients}
Firstly, the local interstellar medium (LISM) is known to be a highly
asymmetric low density region, extending between 65 to 250~pc 
-- recent studies have for the first  
time conducted tomography of the LISM~\citep{2003A&A...411..447L} --.
As a consequence, the spallation rate is lessened so that
the primary fluxes are expected to be enhanced.

Secondly, the history of the origin of the local bubble
(see \cite{2004astroph} for a review) is an imprint of the explosive
stellar activity in the Solar neighbourhood~\citep{2001ApJ...560L..83M,2002A&A...390..299B,2002PhRvL..88h1101B}.
The spatio-temporal position of these sources is certainly of importance.
However, as the steady state is assumed in this paper, only the
spatial influence can be inspected. For example,
\citet{1979Ap&SS..63...35L,1993ApJ...402..185W} have shown
that an absence of nearby sources (in a diffusion model) can truncate 
the path lengths in weighted slab models. This
truncation has been recognised to be of great importance
in the propagation of heavy nuclei~\citep{1993ApJ...403..644C,1996ApJ...470.1218W}.
This effect is realized here as a circular hole of a few hundreds
of pc surrounding the Solar area reflecting the fact that no
very recent sources have exploded in the Solar neighbourhood.
Note that this truncation could also be an effect of
the matter traversed during the acceleration and the escape from
a source region.

\subsection{ The model }
\label{cylmodel}
\begin{figure}[t]
\begin{center}
\includegraphics[width=5cm]{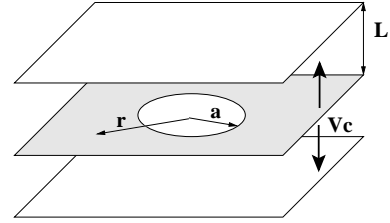}
\caption{Schematic view of the cylindrical geometry for the model used 
in this paper: the hole region of radius $a$ is located in a thin disk
surrounded by a diffusive halo of size $L$ (a convective wind $V_c$ 
may be present).
The exact radius of the Galaxy radius $R$ is not important (see text). 
\label{schematoy}}
\end{center}
\end{figure}

The effects of a local sub-density on radioactive species have been studied 
in the context of a diffusion model by \citet{2002A&A...381..539D}. 
As it is difficult to derive an analytical solution of
the diffusion equation using
a realistic distribution for the gas, it is assumed
that the LISM is a circular cavity of radius $a$ (see below).
It appeared that the radioactive flux
received on Earth was lessened by $\sim$35\% when 
the decay length of the radioactive species\footnote{In a diffusion model,
the  typical length for decay is defined as $l_{\rm rad} \equiv \sqrt{K\gamma\tau_0}$
where $K$ is the diffusion coefficient, $\gamma$ the Lorentz factor and $\tau_0$ the 
lifetime of the nucleus.} 
was twice the size
of the sub-density compared to the case when a sub-density was not
taken into account (cf. \cite{2002A&A...381..539D}, Fig. 4).  
Furthermore, in order to match measured radioactive abundance ratios 
($^{10}\rm{Be}/^{9}\rm{Be}$, $^{26}\rm{Al}/^{27}\rm{Al}$ and
$^{36}\rm{Cl}/\rm{Cl}$), the size
of the local sub-density was constrained to lie within 60 and 100 pc -- 
values consistent with direct observations. This increases confidence in
the model. In the latter, the Galaxy is considered
as a thin disk where the sources and gas are located
(Fig. \ref{schematoy}). To be consistent with
the results obtained for light stable nuclei~\citep{2001ApJ...555..585M}, for
radioactive nuclei~\citep{2002A&A...381..539D} and also
for antiprotons~\citep{2001ApJ...563..172D}, the same geometry
is used. The width of the disk, $2h$, is 200~pc and 
the Galactic radius $R$ is 20~kpc. The exact value
for $R$ is not important so that one can set the center of the
cylindrical geometry at the Earth position (see~\cite{2002A&A...381..539D}),
which simplifies the calculations. The diffusive halo, empty of gas and sources,
extends to a height $L$ on each side of the disk which can vary from 1 to
10~kpc. The local sub-density is then a hole of radius $a$ in this disk,
whereas the source sub-density is a hole of radius $b$.
From the disk emerges a galactic wind with a constant velocity $V_c$ that
adds a convective component to the diffusive propagation of the CRs.

These assumptions (sources and gas in a thin disk, circular hole,
halo empty of gas, same diffusion coefficients in the disk and in the halo) are
probably too strong. However, they have lead to a successful description of
many species in a consistant way, an issue we think to be of great importance.
We also believe that the approximations made (some that would need a
heavy numerical treatment to be relaxed) would have a minor effect
compared to the one induced by the LISM geometry. 
Despite many weaknesses, the present diffusion model has to be 
thought as a first step towards a more realistic description.

Energy losses are not taken into account meaning that in this cylindrical 
geometry the diffusion equation reads
\begin{equation}
-K\Delta N+V_c\frac{\partial N}{\partial z}
+2h\Gamma \delta(z)N=q(r,z)\delta(z)
\end{equation}
where $N=N(r,z)$ is the differential density in energy.
The Dirac distribution $\delta (z)$ expresses
the fact that the sources and the gas are only located in the thin disk.
In this work, the results are always normalized to the source spectrum
thus the only quantity one must specify is $q(r)$, 
the radial source distribution (cf.~\ref{source}).
It is appropriate to use a decomposition in 
Bessel space to solve this equation and the complete derivation 
is given in detail in Appendix~\ref{appencylin} (the derivation closely follows 
that given in~\cite{2002A&A...381..539D}).

The height $L$ of the halo is
a free parameter as are the galactic wind velocity $V_c$, 
the spectral index $\delta$, and normalization $K_0$ of the diffusion
coefficient. Those parameters must be tuned to fit the data.
In the following work and when mentioned, 
the max, median and min set of parameters can be understood 
as detailed in Table \ref{param}.
These three sets are the two extreme and median sets (with regard to 
the value of the diffusion coefficient normalization)
 shown to be compatible with B/C analysis.
Eventually, throughout the paper, the gas sub-density is set to $a=100$~pc whereas
the sub-density in sources can be varied.
\begin{table}[ht]
\centering
\begin{tabular}{c c c  c c}
\hline\hline 
set & $\delta$ & $K_0$ (kpc$^2$~Myr${^{-1}}$) & $L$ (kpc) & $V_c$ (km~s$^{-1}$) \\ \hline 
max & 0.46 & 0.0765 & 15 & 5\\
med & 0.7 & 0.0112 & 4 & 12 \\
min & 0.85  & 0.0016 & 1 & 13.5  \\
\hline
\end{tabular}
\caption{Two extreme and one median sets of parameters shown compatible
with B/C analysis \citep{2001ApJ...555..585M}. \label{param}}
\end{table}

In the remainder of this section, 
we focus on the differences between a model with
a hole in the surrounding gas (and/or sources) and the standard diffusion model with
no hole. Note that as the propagation parameters
used for the standard diffusion model (with no hole) are fitted to B/C,
the results obtained in this latter model and those obtained in a LB
are similar (see Fig.~\ref{propCorrFactor}). 
It is useful to define the ``enhancement factor" which is defined as 
the ratio of a given hole configuration to the standard diffusion 
model (no hole, hereafter SDM) for a given set of propagation parameters.
Such a ratio ensures that the source spectral shape and normalization
are factored out for primaries and the production cross section is
for secondaries.
The discussion of the absolute effects on abundances --~including the
standard result from the leaky box model~-- is contained in
Sect.~\ref{Sec:Data}. 

\subsection{General behaviour}

To separate the effects of the absence of gas and sources in the
hole, we study three different configurations: 
\emph{i}) in the first case, 
only the gas is absent from the ``hole region'' but the sources are still 
present, \emph{ii}) the second case is the opposite configuration and the
hole is a sub-density of sources but not of gas 
and \emph{iii}) finally, we consider the situation where there is
both a hole in gas and a hole in sources (possibly having different sizes).
\subsubsection{Profiles}
\begin{figure}[th]
\begin{center}
\includegraphics[bb=12 50  730 580, width=8.8 cm]{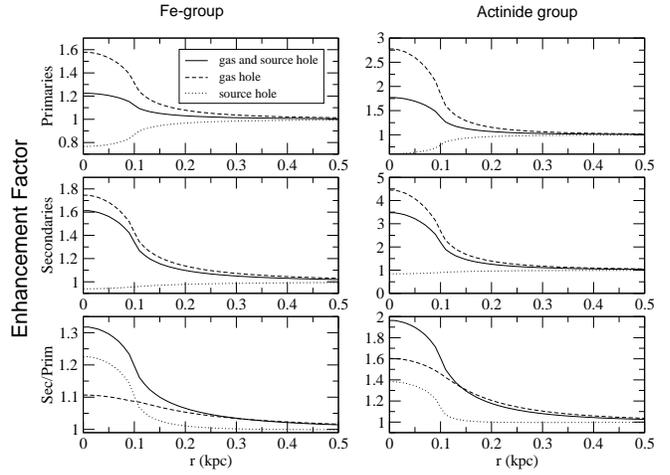}
\caption{Left panels: Fe-group. Right panels: Actinide group. 
Primaries (upper), secondaries (middle) and sec/prim (lower) enhancement
factors are plotted as
a function of the distance to the center of the sub-density
(the fluxes are normalized relatively to the SDM fluxes).
In both models, the hole in gas and the hole in sources are equals
$a=b=0.1$~kpc and $E_k$/n=1~GeV/n. The set of astrophysical parameters is
the median set: $K_0=0.0112$~kpc$^2$~Myr$^{-1}$, 
$L=4$ kpc, $V_c=12$~km~s$^{-1}$ and
$\delta=0.7$\label{densityprofile}. 
}
\end{center}
\end{figure}
The enhancement factors for primaries, secondaries and for the
ratios Secondary/Primary are plotted on Fig.~\ref{densityprofile} as 
a function of the distance to the center $r$ 
($r=0$ corresponds to Earth location). The three hole configurations
described above have been studied. For these profiles, 
the size of the hole is 0.1 kpc, and the median set of parameters
has been used, namely: $K_0=0.0112$~kpc$^2$~Myr$^{-1}$, $L=4$~kpc, 
$V_c=12$~km~s$^{-1}$ and $\delta=0.7$. The iron group (left panel) and
the actinide group (right panel) are both considered at 1~GeV/n.  
Firstly, it appears that when the distance from the hole is 
large enough (typically $3 a$), the enhancement factor
tends to unity (i.e. there is no difference from the no-hole case) 
emphasing that, in a first approximation, 
the hole has only a local effect. An important consequence
is that nuclei are not sensitive to other density features of the ISM, which
justifies the crude model used here.

When a hole in gas only is considered, more primaries (upper panels) 
are expected
(i.e., enhancement factor $> 1$) compared to the SDM as nuclei
entering the hole region do not undergo spallation as they would
if gas were present. In that case, at the Earth location, 
there is a 60\% increase in primaries
of the iron group and a 275\% increase in the actinides.
In the opposite configuration (sources absent, gas present), no primary
nuclei are produced within the sub-density and those which have propagated
to the hole undergo spallation. In that case, the primary density is lessened 
compared to the SDM ($\sim$~20\% for Fe-group nuclei and 
$\sim$~40\% for actinides). 

The third case, where both gas and sources
are absent from the hole region is found to give intermediate results. 
It appears that in this
geometry the absence of gas has a larger impact that the absence of 
sources: for a hole in gas and sources, the enhancement factor 
is greater than unity. There is a 20\% effect for nuclei of the iron group
and a 75\% effect for actinides. 
However, the result is quite sensitive to the particular geometry assumed. 
To obtain a very rough estimate of
this sensitivity, the diffusion equation was solved
for a spherical geometry where sources and gas holes were taken as shells.
In this latter configuration (which is unrealistic, but which may be
considered as the extreme opposite geometry to that of the thin disk case)
the enhancement and relative importance of a hole in gas or a hole in sources are
smaller and the gas-hole enhancement is almost cancelled by a hole in sources. 

When looking at the secondary enhancement factor 
(Fig.~\ref{densityprofile}, middle panels), 
a sub-density in sources only 
has almost no consequences ($\sim$ 10\% effect) on the secondary flux 
(enhancement factor~$\sim$ 1). This indicates that most of the secondaries
found in the solar neighbourhood originate from primaries that 
were not produced locally. When a gas hole is considered (close sources
are present), the secondary enhancement factor is maximal 
($\sim 80$\% for Fe-group nuclei and $\sim 450$ \% for actinides). In this case,
the secondaries present at $r=0$ cannot have been produced locally (as
there is no gas) and have all been propagated from further regions.
Once they reach the gas sub-density, they cannot be destroyed, which
explains the enhancement. The enhancement in the case of a 
hole in gas and sources is explained in the same manner. However, this enhancement is 
slightly lower ($\sim 60$\% for Fe-group nuclei and $\sim 350$\% for actinides)
than the case in which only gas is excluded: when sources
are present in the hole, primaries from the hole can propagate in the 
disk and produce secondaries that may diffuse back towards the Earth.
\subsubsection{Spectra}
\begin{figure}[th]
\begin{center}
\includegraphics[bb=12 50  730 580, width=8.8 cm]{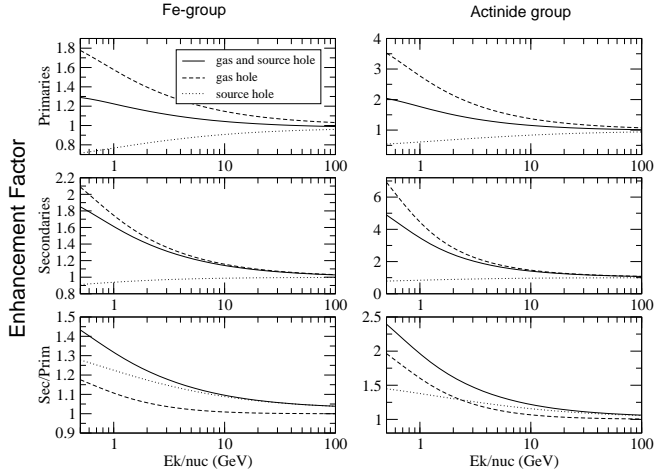}
\caption{Same as Fig.~\ref{densityprofile} but 
for spectra at $r=0$.\label{spectra} 
}
\end{center}
\end{figure}

Enhancement factor spectra for primaries,
secondaries (Fe-group nuclei in the left panel and actinides in the right panel) 
and for the ratio Sec/Prim are plotted in Fig.~\ref{spectra}
between 500~MeV/n and 100~GeV/n at $r=0$.
At high energy the behaviour of any hole configuration tends,
as expected, to the no-hole case at high
energies as spallation becomes negligible. In the case of a sub-density in gas,
the enhancement factor for the median propagation configuration
is quite large (almost a factor of 2), even for the Fe flux. Comparatively,
the sub-Fe/Fe ratio is enhanced by a mere 20\%.

Actually, these enhancements, whether for primaries or secondaries, are
not very sensitive to the exact value of $L$ and $V_c$ (because the
enhancements are normalized to the SDM). They mostly depend
on the value of the diffusion coefficient. Hence, it implies, for example, 
that a very
precise measurement of the iron abundance spectrum would give some hint as to the
value of the diffusion coefficient, by evaluating the deviation from
a standard diffusion model. The larger the deviation at low energy,
the smaller the diffusion coefficient required to fit the deviation.

\subsection{Other dependences}

\subsubsection{Enhancement factor {\em vs} $r_{\rm sp}$}
\label{subsection:cylindrical}
\begin{figure}[th]
\begin{center}
\includegraphics[width=8.5cm]{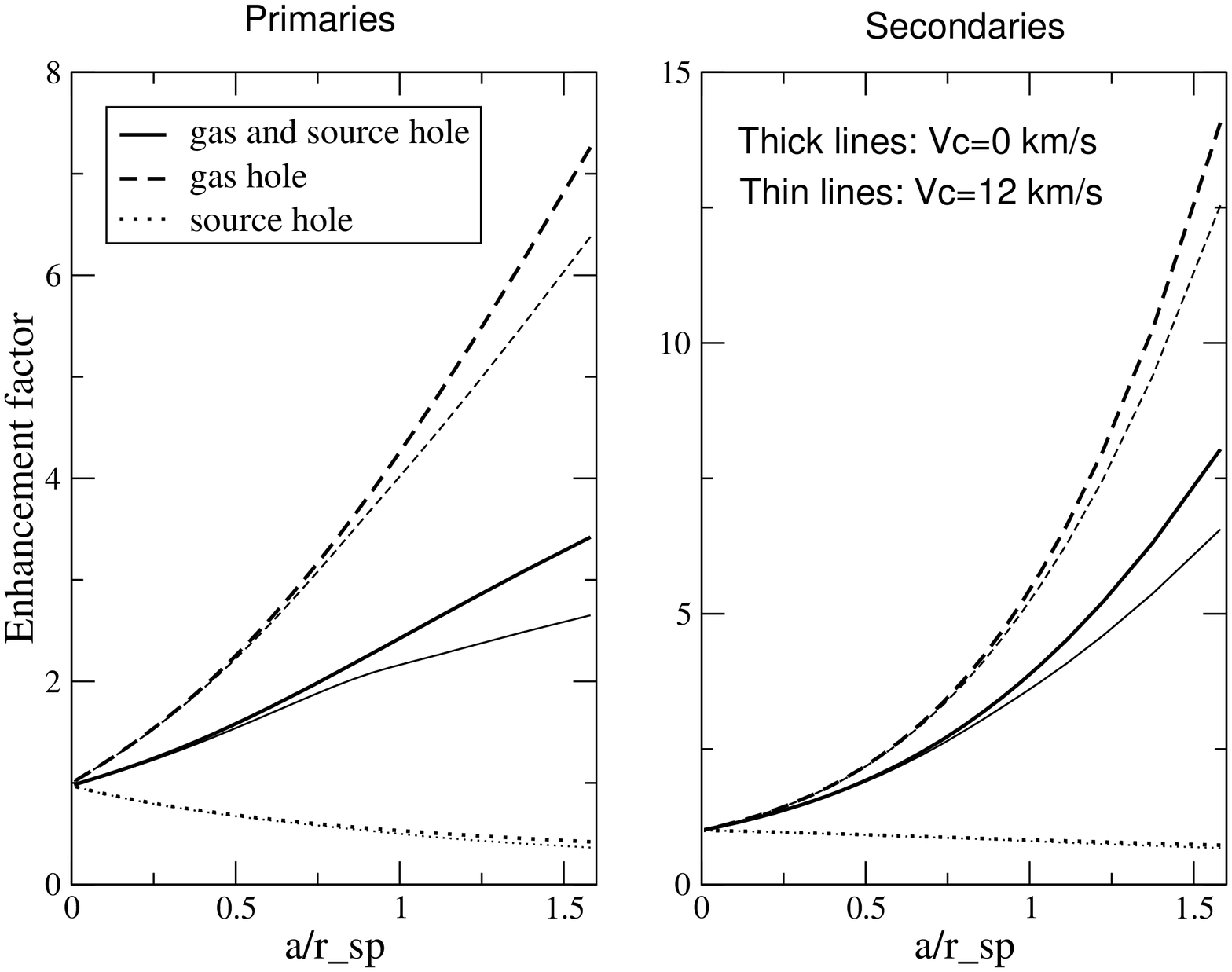}
\caption{Primary and secondary enhancement factors as a function of the
ratio $a/r_{\rm sp}$
for the median set of parameters.
The size of the holes are $a=b=0.1$ kpc, and the different values of 
$r_{\rm sp}$ cover a wide range of energies, $K_0$ and destruction
cross section values.\label{Enhancement}}
\end{center}
\end{figure}

It is useful now to display the enhancement factor as
a function of $a/r_{\rm sp}$ ($a$ being the size of the hole and $r_{sp}$
the typical spallation distance, see Sect.~\ref{time_scales}). 
As several combinations of
$(K_0$, $\delta)$ for a given choice of $A$ yield the same
$r_{\rm sp}$ \citep{2003A&A...402..971T}, the parameter dependence 
is more economically studied.
Indeed, to quickly obtain the enhancement due to a hole in gas
--~or any combination of holes, compared to the SDM~-- 
it is sufficient to take the desired $K_0$ and
$\delta$, choose a specific $A$ and evaluate $r_{\rm sp}$ for a given
energy. The enhancement factor is then directly inferred from
Fig.~\ref{Enhancement}. Two cases are displayed to emphasize
the effect of the Galactic wind: the first with $V_c=0$~km~s$^{-1}$
and the second with $V_c=12$~km~s$^{-1}$. This is the typical range
of the possible values for the wind. It is not a dominant effect.

The enhancement factor for several nuclei could be
obtained from the previous plot, but for simplicity,
they have been gathered
for light to heavy nuclei for several energies and propagation sets 
in Table~\ref{enhancement_factor}. Despite the fact that the three
sets of propagation parameters have been shown to be compatible with B/C analysis,
the minimal set gives some unrealistic enhancement factors. 
This allows the exclusion of an unrealistic configuration for
which the diffusion slope required to fit the B/C data was 
0.85~\citep{2002A&A...394.1039M}, far away from any theoretical expectations. 
Actually, in previous papers attempting to obtain some conservative value
for the propagation parameters from the sole B/C and sub-Fe/Fe
ratios~\citep{2001ApJ...555..585M,2002A&A...394.1039M}, the hole configuration
has not been taken into account. In order to be able to firmly exclude
 some $K_0$ and $\delta$ values, a new detailed analysis is required,
which goes beyond the scope of this paper. In conclusion,
it seems that very small values of $K_0$ have to be excluded because of the
too large enhancement they yield when a hole in gas is included.

\begin{table}[ht]
\centering
\begin{tabular}{c c c}
\hline\hline 
			  & Prim	   &   Sec	   \\
		    & min/med/max    & min/med/max    \\\hline 
LiBeB-CNO      &		&		 \\
E=1~GeV/n~~~~  &  2.08/1.17/1.03  &  2.29/1.13/1.02   \\
E=5~GeV/n~~~~  &  1.36/1.07/1.02  &  1.33/1.05/1.01   \\
E=10~GeV/n~~~  &  1.21/1.04/1.02  &  1.19/1.03/1.01   \\\hline 
Fe-group       &		&		 \\
E=1~GeV/n~~~~  &   6.24/1.58/1.11  &  20.9/1.75/1.11  \\
E=5~GeV/n~~~~  &   2.34/1.23/1.06  &  3.38/1.26/1.06  \\
E=10~GeV/n~~~  &   1.71/1.15/1.04  &  2.03/1.16/1.04  \\\hline
Z=44-48        &		&		 \\
E=1~GeV/n~~~~  &  8.67/1.80/1.14  &  43.0/2.14/1.15   \\
E=5~GeV/n~~~~  &  2.95/1.31/1.08  &  5.06/1.36/1.08   \\
E=10~GeV/n~~~  &  1.99/1.19/1.06  &  2.59/1.22/1.06   \\\hline
Actinides      &		&		 \\
E=1~GeV/n~~~~  &   17.6/2.77/1.26  &  200/4.45/1.31   \\
E=5~GeV/n~~~~  &   5.73/1.61/1.15  &  18.2/1.82/1.16  \\
E=10~GeV/n~~~  &   3.28/1.37/1.11  &  6.22/1.46/1.12  \\
\hline
\end{tabular}
\caption{Enhancement factors at the Earth's location
due to the presence of a local sub-density
on four groups of nuclei. Different energies (from 500~MeV/n to 10~GeV/n)
and three sets of parameters have been considered. \label{enhancement_factor}}
\end{table}

\subsubsection{Impact of the hole in sources $b$}
A few decades ago, some authors introduced the possibility
of a truncation of the path lengths used in 
the weighted slab formalism. \cite{1979Ap&SS..63...35L} then showed
that a hole in sources in a diffusion model mimics such path-length 
truncation. This truncation was first introduced
to fit best the sub-Fe/Fe ratio, though afterwards, new results
on cross section production reduce the need for it. 
However, truncation was implemented for UHCR
propagation and it has a sizeable
effect~\citep{1993ApJ...403..644C,1996ApJ...470.1218W}, especially
for secondary nucleus production. The truncation uncertainty is here
equivalent to the uncertainty in $b$ (see also
Fig.~\ref{propCorrFactor} in Sec.~\ref{sec:Abund_norm_bias}).

\begin{figure}[ht!]
\begin{center}
\includegraphics[bb=12 50  730 580, width=8.8 cm]{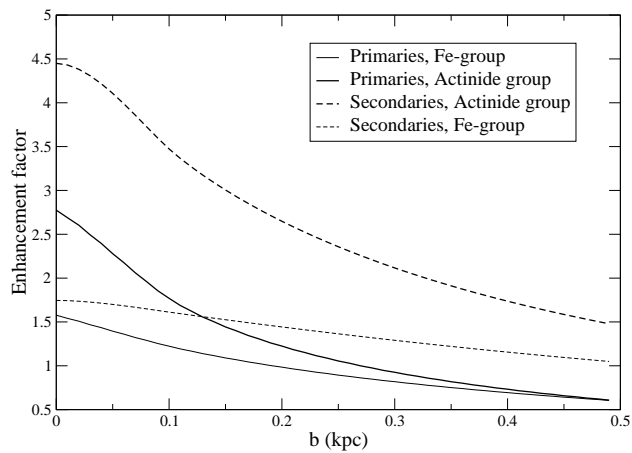}
\caption{ 
Enhancement factor for primaries and secondaries of the Fe and 
actinide groups as a function of the size of the hole in sources,
$b$. A constant gas hole radius $a=100$~pc was considered and
the propagation computed using the median set of parameters.
\label{asrc} 
}
\end{center}
\end{figure}

In this section, the hole in sources varies from $b=0$ to $b=500$~pc.
In Fig.~\ref{asrc}, the enhancement factors of primaries and secondaries
of the iron group and actinide group are plotted as function
of the size of the hole in sources. The energy
of the nuclei is 1~GeV/n and the size of the gas hole is once again 
fixed at $a=100$~pc
(because this feature is certainly present).
Depending on the hole size $b$, one can have 
an enhancement factor that is greater or lower
than unity. Actually, this situation occurs
only because the hole in sources is in competition with the hole in gas.

However, we do not wish to put too much emphasis on this
effect. It must be considered as an illustration of the
impact on relaxing the condition of continuous distribution of sources.
The limits of our model are reached and to go further, one must
relax the steady state approximation. In fact,
the question of nearby sources cannot be easily disconnected to
the question of recent sources \citep{2004ApJ...609..173T}.

\subsection{Preliminary conclusions}

Setting holes in a diffusion model affects the abundance spectra, especially
at low energies where the importance of spallation is larger. A hole in
gas increases the number of primaries, even more the number of secondaries.
Conversely, a hole in sources decreases the number of primaries (the secondaries
being quite insensitive to it). So there is a balance between the
two configurations. Unless a large hole in sources is chosen ($b\gtrsim300$~pc),
because the hole in gas (size $a\sim100$~pc) is firmly established --
from direct observations, and also indirectly because it better fits light
radioactive CRs measurements~\citep{2002A&A...381..539D} --, the final effect is an enhancement
of the cosmic ray fluxes. 

The dependence of this enhancement to the halo
size $L$ and the convective wind $V_c$ is minor. The main dependence
is through the diffusion coefficient, $K(E)$. The previous sections helped
us to establish that very small diffusion coefficients that
were found to best fit the B/C ratio in a standard diffusion
model~\citep{2002A&A...394.1039M} must be discarded as
they produce a too large enhancement of fluxes when the hole in gas
is taken into account. It is reassuring since these small values for
$K_0$ corresponded to quite large diffusion slopes, unsupported by
theoretical considerations (too far from a Kolmogorov or a Kraichnan
turbulence spectrum). However, a new study of B/C must be conducted taking
into account the hole configuration to provide a quantitative result.
Going one step further, one sees that the low energy secondary fluxes
can be useful to determine $K_0$. The secondaries better suit this 
estimation than primaries
as, unlike the latter, they are not very sensitive to a hole in sources.
 This could provide another way,
apart from using radioactive nuclei \citep{1998A&A...337..859P}, to extract the 
diffusion coefficient without too much
ambiguity. Note also that the heavier the
nucleus, the more local it is, so that looking at different nuclei gives
different sampling regions where the diffusion coefficient is averaged.  
Figure~\ref{CorrUHCRs} in the next section will provide a better understanding
of how the effect of the hole sometimes disconnects from the dependence on $L$
and so in certain configuration why the flux at low energy only depends on
$K_0$ (if the latter is not too large).

One can now come to the differences in derived source abundances. Cosmic ray
fluxes are propagated to source by using a leaky box 
(or equivalently a standard
diffusion model) and hole models applied to real data.

\section{Consequences for CRs observations}
\label{Sec:Data}
UHCR abundance spectra have been obtained via spacecraft measurements 
since the 1970s, most notably by {\sc ariel}~6~\citep{1987ApJ...314..739F},
{\sc heao}-3~\citep{1989ApJ...346..997B}, {\sc trek} \citep{westphal} and the
Ultra-Heavy Cosmic-Ray Experiment (or {\sc uhcre}; 
Donnelly et al., in preparation). 

The data are scarce, especially in the actinide region, and only 
elemental abundances
have been obtained. However, several important conclusions have 
already been drawn.
They are
related i) to nucleosynthesic aspects (see the introduction), ii) to
the possible site of acceleration and iii) to the mechanisms leading to
elemental segregation during acceleration.

The recent results of {\sc uhcre} provide unprecedented statistics
and will allow to give firmer conclusions. However, as emphasized
throughout this paper, some aspects of the propagation still need to 
be clarified
in order to fully interpret these data. In this last section,
some questions and results about UHCRs are first reiterated, before inspecting
the propagation effects on these nuclei in different models. In particular,
those species suffering from major propagation uncertainties 
(regardless of cross-sections, which is another issue) are
sorted.

\subsection{Introduction: UHCR data, their interpretation and general issues}

The ultra-heavy ($Z > 70$) elemental abundance ratios most 
pertinent in determining cosmic ray origin are Pb-group/Pt-group, 
$_{92}$U/$_{90}$Th and Actinides/Pt-group\footnote{Pt-group$\equiv(73.5
\leq Z \leq 80.5)$, Pb-group$\equiv(80.5 \leq Z \leq 83.5)$ and 
Actinides are $Z \geq 88$.}.

The first, Pb/Pt, provides clues as to biases in the CR-source 
abundances. Elements in the Pt-group are mainly intermediate-FIP, 
refractory and r-process, while in the Pb-group low-FIP, volatile 
and primarily s-process elements predominate. Thus an anomalous 
GCR Pb/Pt ratio (relative to Solar values) would show up any 
nucleosynthetic (s- or r-process) or atomic (FIP or volatility) bias 
in the source abundances. 
This ratio is more enlightening than 
(for example) the Pt/Fe or Pb/Fe ones, as the ratios of nuclei similar
in mass, are supposed to be relatively 
unaffected by interstellar propagation. This point is discussed in the
next section.
In common with other space-based measurements, the {\sc uhcre} results
(Donnelly et al., in preparation)
demonstrate that the Pb/Pt abundance ratio is decidedly low 
in the GCRs ($0.24 \pm 0.03$) compared to the best estimates from solar 
and meteoritic material ($1.03 \pm 0.12$; \cite{2003ApJ...591.1220L}). Even assuming a 
very severe propagation effect on this ratio ($\times 2.6$), the GCR value is 
a mere $0.63 \pm 0.09$. This could indicate a volatility-based 
acceleration bias as Pb elements are relatively volatile 
\citep{1997ApJ...487..197E,1998SSRv...86..179M}.

The relative abundance of chronometric pairs such as $_{90}$Th and $_{92}$U 
can provide an estimate of the time elapsed since their nucleosynthesis. 
Most models of actinide decay indicate that $_{92}$U/$_{90}$Th drops to unity 
about 1 Gyr after nucleosynthesis.
However, only 44 actinides have been detected so far. 35 of these 
were detected by the {\sc uhcre} and this experiment provides the only 
estimate yet of the  $_{92}$U/$_{90}$Th in the CRs. 
The 1$\sigma$ upper limit is only $\sim 1$, 
implying that significant $_{92}$U decay has occurred and that the time 
elapsed since nucleosynthesis is relatively large.
The transuranics can also be used as excellent cosmic ray 'clocks' 
since the relative abundances of $_{93}$Np, $_{94}$Pu and $_{96}$Cm 
fall drastically 
107 yr after nucleosynthesis \citep{1974Ap&SS..30..275B}.
Again, the best data available are from the {\sc uhcre}, which detected one 
$_{96}$Cm event. The longest-lived isotope of Cm has a half-life of only 16 Myr. 
This fact, combined with the $_{92}$U/$_{90}$Th 
results from the same experiment 
suggests that the CR-source material is an admixture of old and 
freshly-nucleosynthesised matter, such as that found in superbubbles.

Finally, anomalies in the Actinides/Pt ratio could indicate an 
unusual, possibly freshly-nucleosynthesised component in the cosmic ray 
source matter. Results from all of the space-based measurements indicate 
a high value relative to solar system material, though the uncertainties 
on the latter are large.
The {\sc uhcre}'s Actinide/Pt 
ratio\footnote{i.e. ($Z \geq 88$)/($75 \leq Z \leq 79$).}
 ($0.028^{+0.006}_{-0.005}$) 
in broad agreement with other observations, is higher than in the 
present interstellar medium ($0.014\pm 0.002$) and similar to that of the
protosolar medium ($\sim 0.023$) and the interior of superbubbles 
($0.029 \pm 0.005$). 
These observational values are unadjusted for propagation and so are 
therefore best considered as lower-limits.

There are large uncertainties on the effects of propagation on 
these ratios. For example, estimates of the effects of propagation 
on Pb/Pt vary from factors as low as $\times 1.3$ to as high as 
$\times 2.6$. 
Accurate estimates of propagation effects are therefore crucial to 
interpret the data.

\subsection{UHCR abundances and the correction factor for the propagation}

\begin{figure}[ht!]
\begin{center}
\includegraphics[bb=12 50  730 580, width=8.8cm]{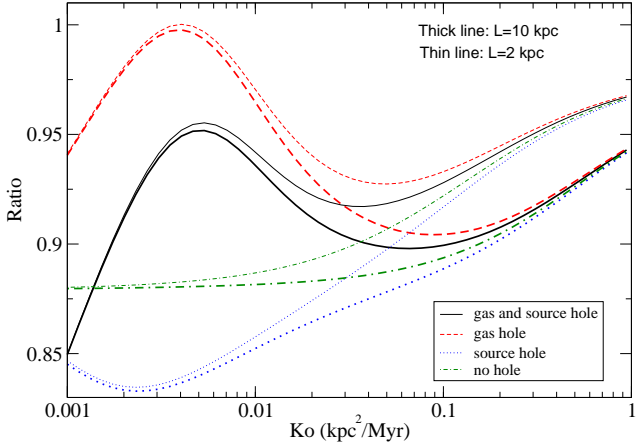}
\caption{${\rm \{Actinides/Pt\}}_{\rm propag.}/{\rm \{Actinides/Pt\}}_{\rm 
source}$  ratio (all species considered as pure primaries). For large diffusion coefficients, the effect of the hole
(any configuration) is negligible and only depends on the halo size 
$L$: the curves with the same $L$
tend to the same values -- upper (resp. lower) group of curves: 
$L=2$ kpc (resp. $L=10$ kpc). On the other hand, for low diffusion
coefficients, the ratio is completely insensitive to $L$.  
See text for details.  \label{CorrUHCRs}
}
\end{center}
\end{figure}

As emphasized above, in the context of UHCRs, one issue is
the determination of relative source abundances of, for example, 
Actinides/Pt and Pb/Pt.
To this end, in this subsection, it is verified that the choice of a 
more refined propagation model leaves the
abundance ratio of close heavy nuclei (if considered as being 
pure primaries) almost unchanged.
Figure~\ref{CorrUHCRs} shows that for any hole configuration and any
propagation parameters, the maximum effect on the 
Actinides/Pt ratio (when both considered as pure primaries) 
is at most 20\% (as obtained in the source hole case).
The effect is even smaller for the Pb/Pt ratio (also considered as 
pure primaries). However, Pt is a mixed species and is responsible for 
the uncertainties on these two
ratios.  
Considering for example the ratio 
$\rm {Pt}^{sec}/\rm{Pb}^{prim}$, the difference in Pt production 
from Pb in the various hole model can be as large as a factor of 2.
Hence, this hole must be taken into account using a complete propagation
network to derive abundances in a model that is thought to be more
realistic than the LB model.

Figure~\ref{CorrUHCRs} is also useful in understanding 
the propagation regimes of heavy nuclei. 
For very small diffusion coefficients (equivalent to low energies),
the spallation length is much smaller than the halo size. In that case,
the ratio of two close nuclei is independent of $L$ but sensitive
to any hole. Conversely, for large diffusion coefficients 
(equivalent to high energies),
the spallation length is much larger than the hole size, thus, the ratio
is only sensitive to $L$. Depending on the nucleus and energy under consideration,
one can be in an intermediate regime where both the hole and halo influence
the ratio.

\subsection{Abundance normalization bias \em{vs} A }
\label{sec:Abund_norm_bias}

The {\sc ariel}~6 experiment~\citep{1987ApJ...314..739F} measured the abundances of
Fe-group elements so that they provided abundances normalized to Fe. One peculiar
feature is the overabundance of the $44\leq Z\leq 48$, 
$62\leq Z\leq 69$ and $70\leq Z\leq 74$ groups which are presumed to be predominantly
secondary in origin. Actinides were also found to be  overabundant in this experiment.
Later, in \citet{1989ApJ...346..997B},  CR abundances (from {\sc heao~3}) 
were propagated back using a LB model.
This study rises the possibility that a bias
occurs during the process because the propagation correction factor in a LB
may be incorrect.

\begin{figure}[ht!]
\begin{center}
\includegraphics[bb=12 50  730 580, width=8.8cm]{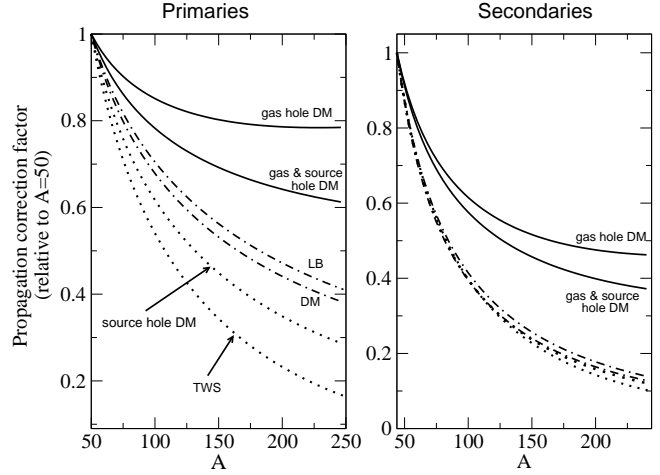}
\caption{Propagation correction factor for a given $A$
relative to the nucleus $A=50$. 
From the curves, one directly gets 
$q_0^{\rm Model~1}/q_0^{\rm Model~2}=
{\rm Corr.Factor}^{\rm Model~2}/{\rm Corr.Factor}^{\rm Model~1}$
which is the relative CR abundances obtained using different propagation
models. 
The standard DM and the LB model  (2 dash-dotted lines) are equivalent.
The hole source DM and the TWS (2 dotted lines) which predict 
a smaller primary density than the LB are also equivalent.
Left panel is for primaries, right panel for secondaries (it is assumed
for simplicity that these secondaries come from the parent $A_{\rm prim}=A_{\rm sec}+6$).
See text for discussion.\label{propCorrFactor}}

\end{center}
\end{figure}
Figure~\ref{propCorrFactor} gives the correction factor according to
the propagation scheme used. As before, for primaries, it corresponds to the
flux divided by the source term whereas for secondaries, it corresponds to
the flux divided by the production cross section. All the results
are normalized to the iron group ($A\sim50$). 
For the diffusion models (without hole and with different
hole configurations), the median set of parameters is used 
($K_0=0.0112$ kpc$^2$~Myr$^{-1}$). We also plotted the results of a
LB model and for a truncated weighted slab (TWS).
The densities in the TWS are calculated 
using the escape mean free path and path length 
distribution (PLD) from \citet{1993ApJ...403..644C} -- detail can be found 
in appendix~\ref{WSappendix}. The truncation was taken to be 
1~g~cm$^{-2}$: \citet{1996ApJ...470.1218W} showed that such a truncation 
combined with a pure $r$-source better fitted the UHCR abundances. 

There is a general trend
showing that the propagation in a LB 
(or as well as in a standard diffusion model) predicts 
 larger fragmentation of nuclei than a diffusion model with a hole in gas,
especially for high A. For a larger diffusion coefficient, this effect would
be weaker. Note that the effect is stronger for secondaries (right panel)
than for primaries (left panel), as already discussed. 
Furthermore, it has to be noticed that both the TWS model
and the source hole diffusion model give lower primary densities that
a LB or a standard DM model. This is pointing towards the study of 
\citet{1979Ap&SS..63...35L} who showed that a truncation of  
the short path lengths
was equivalent to the removal of the nearby sources in a DM. 
As an illustration (not displayed here), we find that to obtain the same curve
with the source hole diffusion model or with the WS model
truncated at $x_0=1$~g~cm$^{-2}$, the size of the hole
has to be set around 180 pc (when using the median diffusion
coefficient). 

\begin{figure}[ht!]
\begin{center}
\includegraphics[bb=12 50  730 580, width=8.8cm]{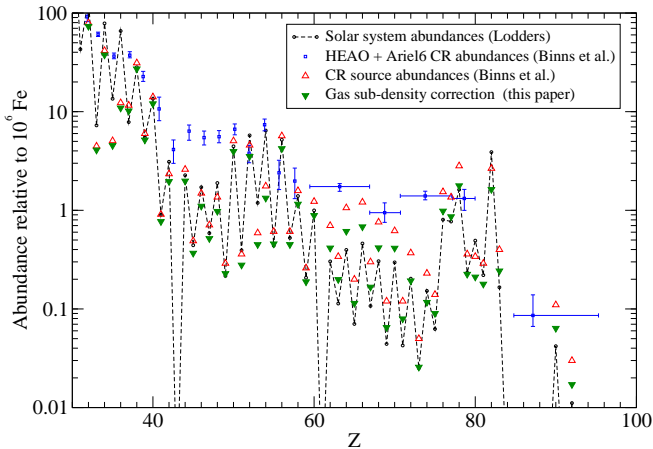}
\caption{Circles: Solar System abundances~\citep{2003ApJ...591.1220L}.
Squares: Measured CR abundances combining HEAO and Ariel 
data~\citep{1989ApJ...346..997B}. The open up triangles represent the CR
source abundances using a Leaky Box model from the above data.
Filled down triangles are these values corrected by the factor
between the LB model to a diffusion model with a hole in gas. These
factors are taken from Fig.~\ref{propCorrFactor}, and correspond to the 
median propagation set. 
For the ranges $44\leq Z\leq 48$, $62\leq Z\leq 69$, $72\leq Z\leq 74$,  
the nuclei were considered to be pure secondaries.
\label{Lodders+Binns}
}
\end{center}
\end{figure}

As the energy of
the events measured in the detectors are not
well known, the propagation correction cannot be easily performed. 
As explained in the caption of the figure, these curves give 
rough corrections to transform LB-derived source abundances to the 
``more realistic" ones in hole models. For illustrative purpose,
we apply this correction to the LB abundances obtained by 
\citet{1989ApJ...346..997B}.
Figure~\ref{Lodders+Binns} shows that for the specific CR abundances obtained
by these authors, the LB-propagated abundances (open triangles) display a 
discrepancy compared to the SS ones above $Z=60$, namely that they are twice larger. 
Previously, a possible enhancement in the r-process contribution 
in this charge range was suggested. 
Applying the approximate correction due to a gas sub-density in our diffusion 
model, a better agreement with SS abundances is obtained (filled triangles)
for heavier species while lighter species are less affected. One must bear 
in mind that this is a very rough estimation.


\section{Conclusion}
\label{Sec:Concl}
The propagation of heavy to UHCRs was considered in a two zone 
diffusion model with various assumptions about the local gas density
and source distribution. The propagation parameters used here
were previously shown to be compatible with B/C and antiproton data, which 
allows a coherent treatment of both heavy and lighter species. Moreover,
a previous analogous study for radioactive species has validated the relevance
of modelling the observed local sub-density as a circular
hole of $\sim 100$~pc in the Galactic disk, increasing confidence 
in the model used here.  

It is well known that for light nuclei, Leaky Box and diffusion 
models are equivalent, which is not the case for the radioactive species.
Using propagation parameters matching B/C ratio in a LB and a DM,
we explicitly checked that the same equivalence exists for the heavy nuclei.
However, this equivalence does not hold when a local sub-density in gas
and/or sources is taken into account.
These new models can be
understood as physical configurations which change
the path length distribution. They are equivalent to truncation 
--~whose importance has previously been underlined for 
the propagation of heavy nuclei~-- for the case of a hole in sources.
To our knowledge, there is no well-know modified PLD corresponding
to the case of a hole in gas.

It was found that for nuclei in the same mass range, these models have a weak 
impact on estimates of CR abundances as long as one deals only
with pure primaries and pure secondaries.
The determination of mixed species abundances, e.g. Pt, is
on the other hand very sensitive to the presence of a local gas
sub-density.
The strength of this effect depends strongly on the diffusion
coefficient value (and hence on the energy of the CRs).

Whereas this effect is small for propagated nuclei of similar masses, 
it becomes 
important when considering the whole range of nuclei from
iron to the actinides. For a typical diffusion coefficient, it
is found that the CR abundances derived in a gas hole model are
lessened by a factor of 2 compared to those evaluated in a LB model. 
In this case, the discrepancy with solar abundances for $Z>60$ nuclei
is smaller. This could have important consequences on the 
of r-process contribution needed to explain the measured data. 
Note that all these results
were derived independently of the value of the production cross sections.

Further work -- using a full propagation network for nuclei -- 
is required to obtain more quantitative results. Before this
work to be completed, a similar study will first be conducted for
nuclei that are unstable to electronic capture: the latter are known
to affect source abundances (see e.g., Table~I in \citet{1985Ap&SS.114..365L})
and the local subdensity is likely to have an effect (slightly different than 
the one observed for $\beta$-unstable nuclei) on their propagation.

\begin{acknowledgements}
This work has benefited from a French-Irish programme, PAI Ulysses, 
EGIDE/Enterprise Ireland.
We thank Dr. Waddington for suggesting the comparison to
the truncated weighted slab model.
\end{acknowledgements}
\appendix
		
\section{Derivation in cylindrical geometry}
\label{appencylin}
The calculation leading to the solutions
for the two zone model in cylindrical geometry without 
energy losses is presented and follows \citet{2002A&A...381..539D}.
The diffusion equation describing the evolution of the 
nucleus density $N(r,z)$, including spallation and convection 
due to the galactic wind reads:
\begin{displaymath}
-K\Delta N(r,z)+V_c\frac{\partial N}{\partial z}
+2h\Gamma \delta(z)N(r,z)=q(r,z)
\end{displaymath}
where $q(r,z)= q(r)\times 2h\delta(z)$ is the source distribution, 
$K$ is the diffusion coefficient, $V_c$ is the velocity of 
the galactic convection wind, $h$ is the thickness of the disk,
and $\Gamma=n \sigma v$ is the spallation reaction rate.
The Dirac distribution is needed as the spallation and
sources are only present in the disk.
Considering a local density with a radius $a$, the density 
is then given by $n=n_{ISM}\Theta(r-a)$, where $\Theta(r-a)$ is
the Heavyside function. 
The space of the Bessel functions is well adapted to this
geometry and we use the following decompositions, with $\zeta_i$,
the $i$-th zero of the Bessel function $J_0$:
 \begin{equation}
N(r,z)=\sum_iN_i(z)J_0\left(\zeta_i\frac{r}{R}\right)
\label{decompbessel}
\end{equation}
\begin{displaymath}
q(r)=\sum_iq_i J_0\left(\zeta_i\frac{r}{R}\right)
\end{displaymath}
\begin{displaymath}
\Theta(r-a)N(r,z)=\sum_i\Omega_iJ_0\left(\zeta_i\frac{r}{R}\right)
\end{displaymath}	
with $\Omega_i$ following
\begin{displaymath}
\Omega_i=
\frac{2}{J_1^2(\zeta_i)}\int_{a/R}^1\rho N(\rho,0)\Theta(\rho-\frac{a}{R})
J_0(\zeta_i\rho)\rm{d}\rho \; .
\end{displaymath}
In the space of Bessel functions, the diffusion equation becomes
\begin{equation}
\frac{\partial^2 N_i}{\partial z^2}
-\frac{V_c}{K}\frac{\partial N_i}{\partial z}
-\frac{\zeta_i^2}{R^2}N_i-\frac{2h\Gamma}{K}\delta(z)\Omega_i=
-2h\frac{q_i}{K}\delta(z) \; .
\label{diffbessel}
\end{equation}
Using the property of the Bessel functions
\begin{displaymath}
\int\rho J_0(\zeta_j \rho)J_0(\zeta_i \rho)\rm{d}\rho=\left\{\begin{array}{ll}
\frac{1}{\zeta_j^2-\zeta_i^2}[-\zeta_i\rho J_0(\zeta_j \rho)J_1(\zeta_i \rho)
+\\ 
\zeta_j\rho J_1(\zeta_j \rho)J_0(\zeta_i \rho)] & \textrm{for $i\neq j$}\\
&\\
\frac{1}{2}\rho^2[J_0^2(\zeta_i \rho)+J_1^2(\zeta_i \rho)] 
& \textrm{for $i=j$}
\end{array}\right.
\end{displaymath}
$\Omega_i$ reads,
\begin{equation}
\Omega_i=
\left\{\begin{array}{ll}
\frac{2}{J_1^2(\zeta_i)}\frac{a}{R}\sum_{j\neq i}M_{ij}N_j 
&\textrm{for $i\neq j$}\\
&\\
 B_iN_i & \textrm{for $i=j$}\\
\end{array}\right.
\label{omegai}
\end{equation}
with $B_i$ and $M_{ij}$ being respectively
\begin{displaymath}
B_i=1-\frac{a^2}{R^2}\frac{1}{J_1^2(\zeta_i)}\left(J_0^2\left(\zeta_i\frac{a}{R}\right)+
J_1^2\left(\zeta_i\frac{a}{R}\right)\right)\; , 
\end{displaymath}
\begin{displaymath}
M_{ij}=\frac{1}{\zeta_j^2-\zeta_i^2}
\left[\zeta_iJ_0\left(\zeta_j\frac{a}{R}\right)
J_1\left(\zeta_i\frac{a}{R}\right)
-\zeta_jJ_1\left(\zeta_j\frac{a}{R}\right)J_0\left(\zeta_i\frac{a}{R}\right)
\right] \; .
\end{displaymath}
Inserting Eq.~(\ref{omegai}) in the diffusion equation, Eq.~(\ref{diffbessel})
becomes
\begin{equation}
\begin{array}{ll}
\displaystyle\frac{\partial^2 N_i}{\partial z^2}
-\frac{V_c}{K}\frac{\partial N_i}{\partial z}-\frac{\zeta_i^2}{R^2}N_i= 
-2h\frac{q_i}{K}\delta(z)& \\
 & \\
\displaystyle +\frac{1}{K}\left(2h\Gamma B_iN_i(0)  
+4h\Gamma \frac{a}{R}\frac{2}{J_1^2(\zeta_i)}
\sum_{j\neq i}M_{ij}N_j(0)\right)\delta(z) \; .
\label{diffbessel2}
\end{array}
\end{equation}
In the halo, the RHS term of Eq.~(\ref{diffbessel2}) is not present. 
Considering the boundary condition $N_i(z=L)=0$, the solution in
the halo is given by 
\begin{equation}
N_i^{\rm{halo}}(z)=\exp\left(\frac{V_cz}{2K}\right)
N_i(0)\frac{\sinh\left(S_i\frac{L-z}{2}\right)}
	{\sinh\left(\frac{S_i L}{2}\right)} \; ,
\label{solhalobis} 
\end{equation}
where $S_i$ is a constant defined as
\begin{displaymath}
S_i=\left(\frac{V_c^2}{K^2}+4\frac{\zeta_i^2}{R^2}\right)^{\frac{1}{2}} \; .
\end{displaymath}
The solution in the thin disk, $N_i(0)$,
 is obtained by integration Eq.~(\ref{diffbessel2})
across the disk between $z=-h$ and $z=+h$ with $h \rightarrow 0$.
Eq.~(\ref{diffbessel2}) becomes
\begin{equation}
\begin{array}{ll}
\displaystyle 2N_i'(0)-2\frac{V_c}{K}N_i(0)+ 2h\frac{q_i}{K} 
-\frac{1}{K}2h\Gamma B_i N_i(0) & \\
 & \\
\displaystyle -\frac{1}{K}
4h\Gamma \frac{a}{R}\frac{2}{J_1^2(\zeta_i)}
\sum_{j\neq i}M_{ij}N_j(0)=0 \; .
\label{intdisk}
\end{array}
\end{equation}
The continuity between the halo and the disk is established
by inserting the halo solution (Eq.~(\ref{solhalobis})) in 
Eq.~(\ref{intdisk}). 
Defining $A_i$ as
\begin{equation}
A_i=V_c+KS_i\coth\left(\frac{S_iL}{2}\right)+2h\Gamma B_i
\end{equation} 
one finds 
\begin{equation}
N_i(0)=\frac{q_i}{A_i}2h\left(1-\frac{4\Gamma a}{q_i R J_1^2(\zeta_i)}
\sum_{j\neq i}M_{ij}N_j(0)\right) \; .
\label{Ni}
\end{equation}
It appears that to calculate $N_i(0)$, one needs to know the
values for all the other orders of the Bessel decomposition. 
We use a perturbative method
to compute these quantities. At the zero-$th$ order, we have
$N_i^{(0)}(0)=2h q_i/A_i$, and calculate recursively the $(n+1)$-th order 
as
\begin{equation}
N_i^{(n+1)}(0)=\frac{q_i}{A_i} 2h\left(1-\frac{4\Gamma a}{q_i R J_1^2(\zeta_i)}
\sum_{j\neq i}M_{ij}N_j^{(n)}(0)\right)
\end{equation}
until the convergence is reached. In practice, the convergence is reached
quite rapidly, after 5 iterations.
Using Eq.~(\ref{decompbessel}), the density of nuclei in
the physical space is then given by
\begin{equation}
N(r,z)=e^{\left(\frac{V_cz}{2K}\right)}\sum_{i=0}^{+\infty}N_i(0)
\frac{\sinh\left(S_i\frac{L-z}{2}\right)}{\sinh\left(\frac{S_iL}{2}\right)}
J_0\left(\zeta_i\frac{r}{R}\right) \; .
\end{equation}
\subsection{Primaries and secondaries}
\label{source}
The previous derivation is valid for any  source term $q(r)$.
For the primaries, we consider the two situations where there
are sources or not in the hole, but in each case, we assume
a constant source distribution $q_0$ with $r$.
In that case, there are analytical expressions for the components
of $q(r)$ in Bessel space, namely
\begin{displaymath}
q(r)=1 \hspace{1.45cm}\rightarrow \;\;\;\;\;
q_i= \frac{2}{\zeta_i J_1(\zeta_i)}
\end{displaymath}
\begin{displaymath}
q(r)=\Theta(r-b) \;\;\;\rightarrow \;\;\;\;\;
q_i=\frac{2 \left(J_1(\zeta_i)-\frac{b}{R} 
J_1(\zeta_i\frac{b}{R})\right)}{\zeta_i J_1^2(\zeta_i)}
\end{displaymath}
The secondaries are only produced by spallation of the primary
nuclei on the ISM, so their source term is $q(r)=\Gamma_{ps}N^p(r)$,
where $\Gamma_{ps}$ is the production reaction of the secondaries
and $N^p(r)$ is the density of primary nuclei.
As a consequence, for the secondaries we compute Eq.~(\ref{Ni}) simply using
$q_i=\Gamma_{ps}N_i^p$
where $N_i^p$, referring to the primaries, have been previously determined.

\section{The Truncated Weighted Slab}
\label{WSappendix}
In this work, we use the truncated weighted slab (TWS) approach as described
in \citet{1993ApJ...403..644C}. The escape mean free path and path length
distribution are defined respectively as,
\begin{displaymath}
\displaystyle\lambda_{esc}=\left\{\begin{array}{ll}
5.5\mathrm{~g~cm}^{-2} & \mathrm {for~R} \leq 7.6 \mathrm{~GV}  \\
&\\
\displaystyle 5.5\times\left(\frac{R}{7.6 {\rm GV}}\right)^{-0.4}  
& \mathrm{for~R~}> 7.6 \mathrm{~GV}
\end{array}\right.
\end{displaymath}
and,
\begin{displaymath}
\displaystyle \mathcal{P}(x)=\left\{\begin{array}{ll}
\displaystyle \frac{1}{\lambda_{\rm esc}} \exp\left[\frac{x_0-x}{x}\right]& 
\mathrm {for~} x\geq x_0 \\
&\\
\displaystyle 0 & {\rm otherwise}
\end{array}\right.
\end{displaymath}
where $R$ is the rigidity and $x_0$ the truncation of the shortest
path lengths ($x_0$=1~g~cm$^{-2}$ in this work). When neglecting the
energy losses and assuming only one primary parent $p$ for a secondary $s$, 
one finds the truncated weighted slab densities of primaries and 
secondaries to be
\begin{displaymath}
\begin{array}{l}
\displaystyle {N_p=\frac{\lambda_p}{\lambda_{esc}+\lambda_p} q_p e^{-x_0/\lambda_p}}
\\
\\
\displaystyle N_s=\frac{\lambda_s \lambda_p}{(\lambda_p-\lambda_s)\lambda_{ps}}q_p
\left[\frac{\lambda_p}{(\lambda_{esc}+\lambda_p)}e^{-x_0/\lambda_p}
- \frac{\lambda_s}{(\lambda_{esc}+\lambda_s)}e^{-x_0/\lambda_s}\right] 
\end{array}\; 
\end{displaymath}
where $\lambda_p$ (resp. $\lambda_s$) corresponds to the mean free path 
of a primary (resp. secondary) with respect to its total destruction
cross section and where $\lambda_{ps}$ is the mean free path of a 
primary relatively to the secondary production cross section.
	
\bibliography{biblio}

\end{document}